\documentclass[fleqn,usenatbib]{mnras}

\usepackage{newtxtext,newtxmath}

\usepackage[T1]{fontenc}
\usepackage{ae,aecompl}

\usepackage{graphicx}	
\usepackage{amsmath}	
\usepackage{amssymb}	

\title[Restarted jet in NGC\,2639]{The Discovery of Secondary Lobes in the Seyfert Galaxy NGC\,2639}

\author[Sebastian et al.]{
Biny Sebastian,$^{1}$\thanks{E-mail: biny@ncra.tifr.res.in}
P. Kharb,$^{1}$
C. P. O' Dea$^{2,3}$
J. F. Gallimore$^{4}$
and S. A. Baum$^{2,5}$
\\
$^{1}$National Centre for Radio Astrophysics (NCRA) - Tata Institute of Fundamental Research (TIFR),\\ S. P. Pune University Campus, Post Bag 3, Ganeshkhind, Pune 411007, India\\
$^{2}$Department of Physics and Astronomy, University of Manitoba, Winnipeg, Canada\\
$^{3}$School of Physics and Astronomy, Rochester Institute of Technology, Rochester, NY 14623, USA\\
$^{4}$Department of Physics and Astronomy, Bucknell University, Lewisburg, PA 17837, USA \\
$^{5}$Carlson Center of Imaging Science, Rochester Institute of Technology, Rochester, NY 14623, USA
}

\begin{document}
\label{firstpage}
\pagerange{\pageref{firstpage}--\pageref{lastpage}}
\maketitle

\begin{abstract}
We report the discovery of a secondary pair of radio lobes in the Seyfert galaxy NGC\,2639 with polarization-sensitive observations with the Karl G. Jansky Very Large Array (VLA). The presence of these lobes, which are aligned nearly perpendicular to the known set of radio lobes observed in the east-west direction, has not been reported previously in the literature. The in-band rotation measure image shows gradients in both the lobes indicative of organised magnetic field structures on kpc-scales. The magnetic field structure is aligned with the jet/lobe direction in both the lobes. Based on the settled optical morphology of the host galaxy, it is likely that a minor merger that did not disrupt the host galaxy structure is responsible for the observed features in NGC\,2639. This also explains the near 90$\degr$ change in the jet direction; the current jet direction being the result of a new accretion disk formed by the minor merger, whose direction was a result of the angular momentum of the inflowing merger gas. 
\end{abstract}


\section{Introduction}
Active galactic nuclei (AGN) are the energetic centres of galaxies that are powered by the release of gravitational potential energy as matter accretes onto supermassive black holes. Major and minor galaxy mergers play a key role in the release of angular momentum of galactic gas, brought all the way from kpc-scales to the small sub-parsec-scale accretion disks, to fuel the AGN \citep{barnes1991,barnes1996}. Several merger events have therefore been suggested to explain the observed recurring AGN activity in many galaxies \citep{liu2003}. 

Among the radio-loud AGN class, double-double radio galaxies are understood to be clear signatures of recurrent AGN activity \citep{schoenmakers2000,sebastian2018}. Interestingly, the different epochs of emission are most often aligned with each other \citep{saikia2006}. The lack of misalignment in most of the sources might point towards a constant black hole spin direction despite the restarting of the jet. Only rarely are some sources possessing misaligned pair of double lobes, seen \citep{akujor1996,nandi2017}. Misalignment in jet directions can be produced by slow precession which in turn can be caused by the warping of accretion disks, binary black holes \citep{begelman1980} or the influence of a nearby neighbouring galaxy \citep{blandford1978}. Coalescence of binary black holes has been suggested to cause a rapid spin flip in the primary black hole producing the jet, thereby producing jet misalignment between epochs \citep{dennett2002}. 

In radio-quiet AGN like Seyfert galaxies however, multiple radio lobes are not often observed. So far, only a handful of Seyfert galaxies have been known to exhibit more than one pair of kpc-scale radio lobes, like Mrk\,6 \citep{Kharb06} and NGC\,2992 \citep{Irwin17}. Here we report the discovery of a second pair of radio lobes  in new 5~GHz polarization-sensitive observations with the Karl G. Jansky Very Large Array (VLA), in the Seyfert galaxy NGC\,2639. The newly discovered north-south oriented lobes are aligned nearly perpendicular to the well-known east-west lobes. NGC\,2639 is a type 1.9 Seyfert (alternately classified as a LINER\footnote{Low Ionization Nuclear Emission Line Region galaxy}) hosted by an SA (ring) type grand-design spiral galaxy \citep{Martini03} at a redshift of 0.011128. At the distance of NGC\,2639, 1$\arcsec$ corresponds to a linear extent of 0.229~kpc for H$_0$ = 73~km~s$^{-1}$~Mpc$^{-1}$, $\Omega_{mat}$ = 0.27, $\Omega_{vac}$ = 0.73. Throughout this paper, spectral index $\alpha$ is defined such that flux density at frequency $\nu$ is $S_\nu\propto\nu^\alpha$.

\section{Observations and Data Analysis} 
\label{sec:obs}
NGC\,2639 was observed using the Expanded VLA (project ID: 17B$\_$074) at 5~GHz (4.424 -6.44 GHz) in full polarization mode using the B array configuration. NGC\,2639 was observed as a part of a larger sample of Seyfert galaxies hosting kiloparsec-scale radio emission (Sebastian et al. 2019, in prep.) The band was divided into 16 spectral windows (spws) each with a bandwidth of 128 MHz. 0832+492 was used as the phase calibrator whereas 3C138 was the flux density as well as polarization calibrator. 3C84 was used for calibrating the polarization leakage terms. Initial data editing and calibration were carried out using the $\tt CASA$ pipeline for EVLA data reduction. The polarization calibration was carried out after the basic flux and phase calibration was completed. We followed the data reduction and imaging steps similar to those elaborated in \cite{sebastian2019}.

\begin{figure}
\centering{\includegraphics[width=7cm]{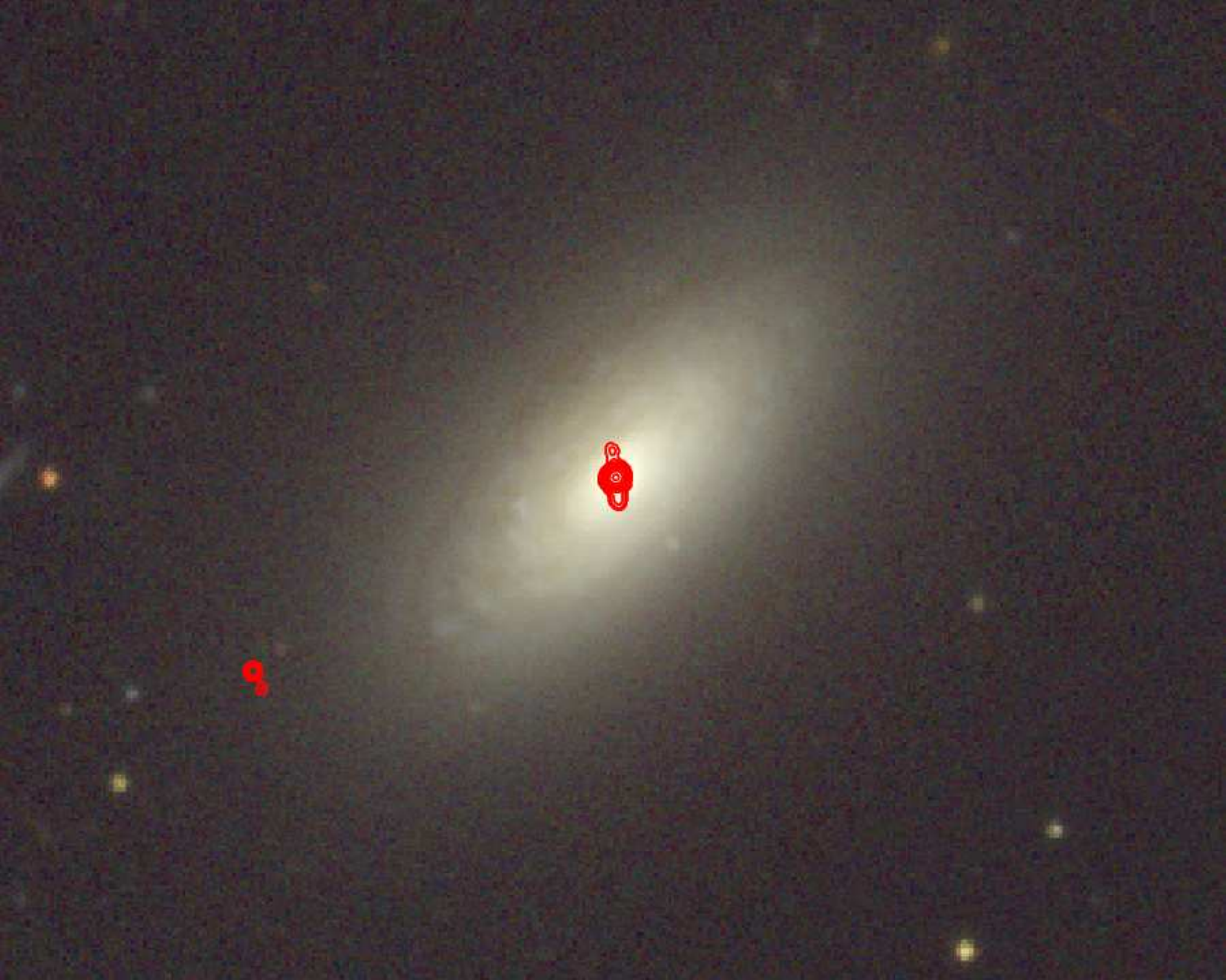}}
\caption{The color SDSS image of NGC\,2639 superimposed by 5~GHz EVLA radio contours in red, clearly showing the presence of the newly discovered north-south radio lobes.}
\label{fig1}
\end{figure}

We made use of the multi term-multi frequency synthesis ($\tt MT-MFS$; see \cite{rau2011} for more details) algorithm while imaging in $\tt CASA$ to correct for wide-band related errors while deconvolving. The I, Q and U images of NGC\,2639 were made using task $\tt tclean$ in CASA. Three rounds of phase-only self-calibration was carried out before a final round of A\&P self-calibration. 
We used nterms =2 while carrying out the $\tt MT-MFS$ clean in CASA. 
In order to obtain accurate spectral index maps, the EVLA calibrated data file was therefore divided into four parts, each consisting of 4 spectral windows. The imaging of these four data sets were carried out using common cell size, image size and a restoring beam (1.22$\arcsec \times $1.22$\arcsec$). We blanked all the pixels which had total intensity values less than 3$\sigma$ in any of the images. We then fitted a power law to flux densities at each pixel as a function of the frequency to estimate the in-band spectral index values.

The total polarization image and the polarization angle images were produced from the Q and U images. We corrected for Ricean bias using the task $\tt COMB$ in Astronomical Image Processing System ($\tt AIPS$). In order to make the rotation measure (RM) image the calibrated data were divided into four chunks each with 500 MHz bandwidth. The images were then convolved to the same beam size of 1.3$\arcsec \times$ 1.3$\arcsec$. The task $\tt rmfit$ was used to generate an RM image from the `QU' Stokes image cubes. The pixels where the error in RM was greater than 150 rad m$^{-2}$ were blanked. The final r.m.s. noise in the EVLA total intensity and polarization images is $\sim10~\mu$Jy~beam$^{-1}$. 
 In addition to the new EVLA data, we also analyzed archival VLA data at 5~GHz from 1998 (Project ID: GL022) and Very Long Baseline Array (VLBA) data at 8.3~GHz from 2011 (Project ID: BC196J). We followed standard data reduction procedures as described in the AIPS cookbook for VLA and VLBA data\footnote{http://www.aips.nrao.edu/CookHTML/CookBookap1.html{\#}x168-369000A.\\ http://www.aips.nrao.edu/CookHTML/CookBookap3.html{\#}x180-386000C.}.

\section{Results}
\label{secresults}
\begin{figure}
\centerline{
\includegraphics[width=10cm]{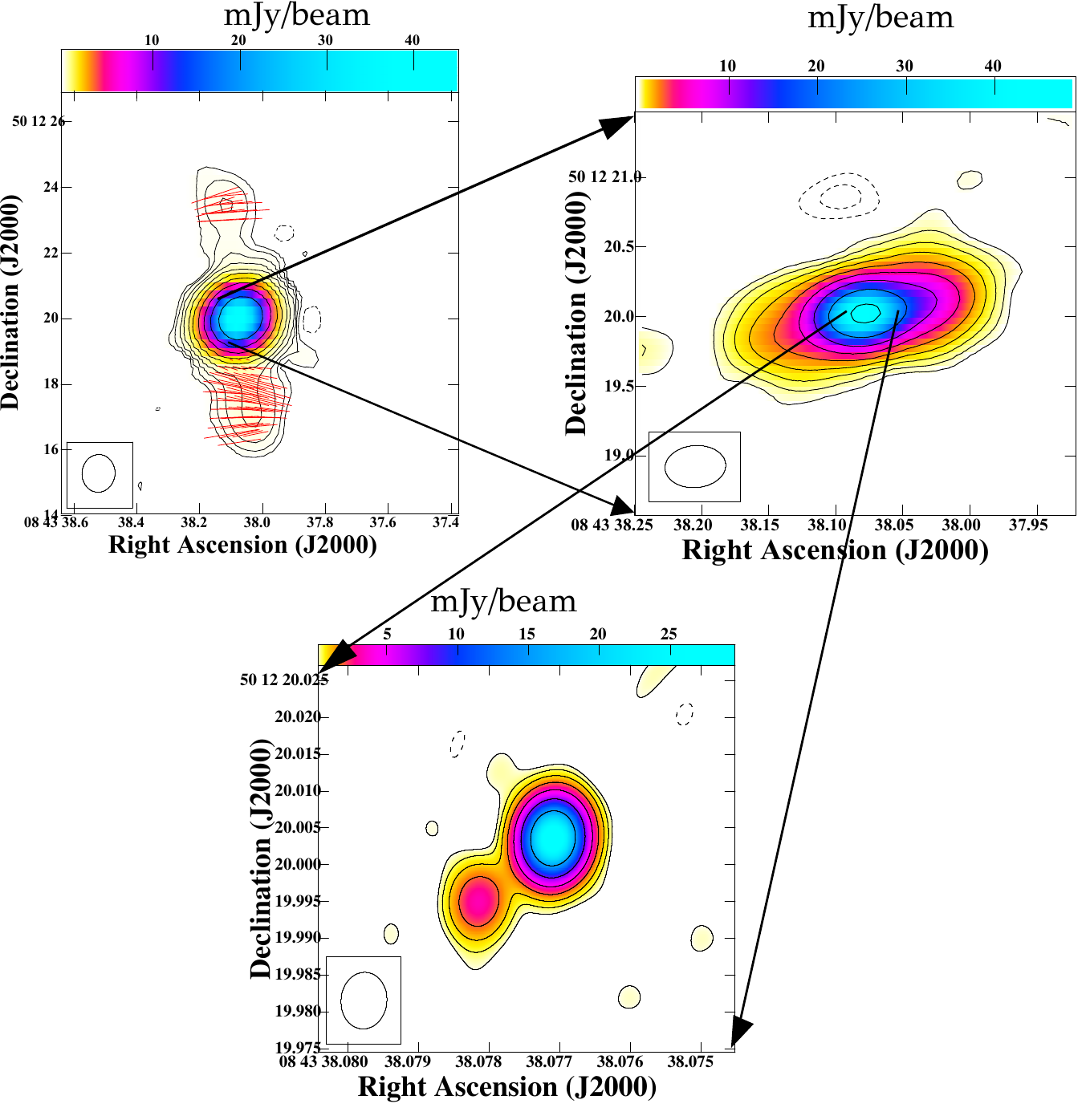}}
\caption{(Top left) 5.5~GHz total intensity radio contours of NGC\,2639 with polarized intensity shown in color. Red ticks denote the polarization electric vectors with lengths proportional to the fractional polarization. Contour levels: 50 $\times$ (-2, -1, 1, 2, 4, 8, 16, 32, 64, 128, 256, 512) $\mu$Jy beam$^{-1}$. The newly discovered north-south lobes are highly linearly polarised. The beam shown in the bottom left corner is of size $1.15\arcsec\times1.00\arcsec$ at PA =$-$4.9$\degr$. (Top right) 5~GHz total intensity radio contours overlaid on colour image of the total intensity of NGC\,2639. Contour levels: 164$\times$($-$2, $-$1, 1, 2, 4, 8, 16, 32, 64, 128) $\mu$Jy beam$^{-1}$. The east-west lobes are embedded within the core of the image in the left panel. The beam shown in the bottom left corner is of size $0.435\arcsec\times0.302\arcsec$ at PA = 85.4$\degr$. (Bottom) 8.3 GHz archival VLBA image showing a $\sim$1~parsec jet at PA = 130$\degr$. Contour levels: 239.2 $\times$ ($-$2, $-$1, 1, 2, 4, 8, 16, 32, 64) $\mu$Jy beam$^{-1}$. The beam shown in the bottom left corner is of size
7.7 mas~$\times$~6.2~mas at PA =$-5\degr$.}
\label{fig2}
\end{figure}

Figure~\ref{fig1} shows the total radio intensity image at 5~GHz superimposed on the SDSS $gri$ band color composite image of NGC\,2639. Figure~\ref{fig2} shows the polarization and total intensity image from EVLA, as well as the total intensity images from VLA and VLBA. The new north-south radio lobes with a prominent `core' are clearly seen in Figures~\ref{fig1} and \ref{fig2}. The `core' is the already known triple core-jet structure observed by \citet{thean2000} and \citet{baldi2018}. The VLBA image shows a one-sided jet which is misaligned with the lobes seen in both the EVLA and the archival VLA images. The linear size of each of the north-south lobes is $\sim$0.8~kpc. The extent of the east-west lobes is $\sim$180~parsec on each side of the core, whereas the VLBA core-jet extent is $\sim$3~parsec. It is clear from the optical image of the host galaxy that the radio lobes are misaligned with the optical disk. The position angle (PA)\footnote{PA is measured from north through east with north being at 0$\degr$} of the host galaxy is $136\degr$ whereas the PA of the north-south radio lobes is $-174\degr$ and that of the east-west lobes is $106\degr$. The radio morphology of NGC\,2639 resembles a Seyfert galaxy with lobes like Mrk\,6 \citep{Kharb06} or NGC\,6764 \citep{Kharb10}. 

The north-south radio lobes are highly polarized showing an average fractional polarization of $20\pm3\%$ in the southern lobe and $30\pm7\%$ in the northern lobe. Although the southern lobe is brighter in total intensity, the average fractional polarization is slightly higher in the northern lobe. For the optically thin radio lobes in NGC\,2639, the inferred magnetic field structures are aligned with the direction of the lobes, but show a slight change in orientation (by about 30$\degr$) at the end of the lobes, both in the north and the south.

The in-band rotation measure (RM) image is shown in Figure~\ref{fig3}. 
The average spectral index in the northern lobes is $\sim-0.26\pm0.05 $ while that in the southern lobes is $\sim-1.4\pm0.3$. The core has a flat mean spectral index of $-0.16\pm0.02$. The RM of the southern lobe has a mean value of $\sim70\pm50$~rad~m$^2$. There appears to be an RM gradient in the southern lobe roughly in the north-west - south-east direction. A suggestion for an RM gradient is also observed in the northern lobe roughly in the east-west direction. The ``minimum energy'' magnetic field strength \citep[see][]{ODea87} for a volume filling factor and proton to electron ratio ($k$) of unity, is 12-14$\mu$G and $\sim88\mu$G for the north-south lobes and `core', respectively, for NGC\,2639.\footnote{For $k=100$, the magnetic field is underestimated by a factor of $(k +1)^{\frac{1}{\alpha+3}}$ or $3-4$.} The volume was estimated using the observed projected length and width of the lobes. The corresponding electron lifetimes \citep[undergoing both synchrotron and inverse Compton losses over CMB photons][]{van69} are 12-16 Myr and 0.8 Myr for the north-south lobes and `core', respectively.

\begin{figure}
\centering{
\includegraphics[width=7.9cm]{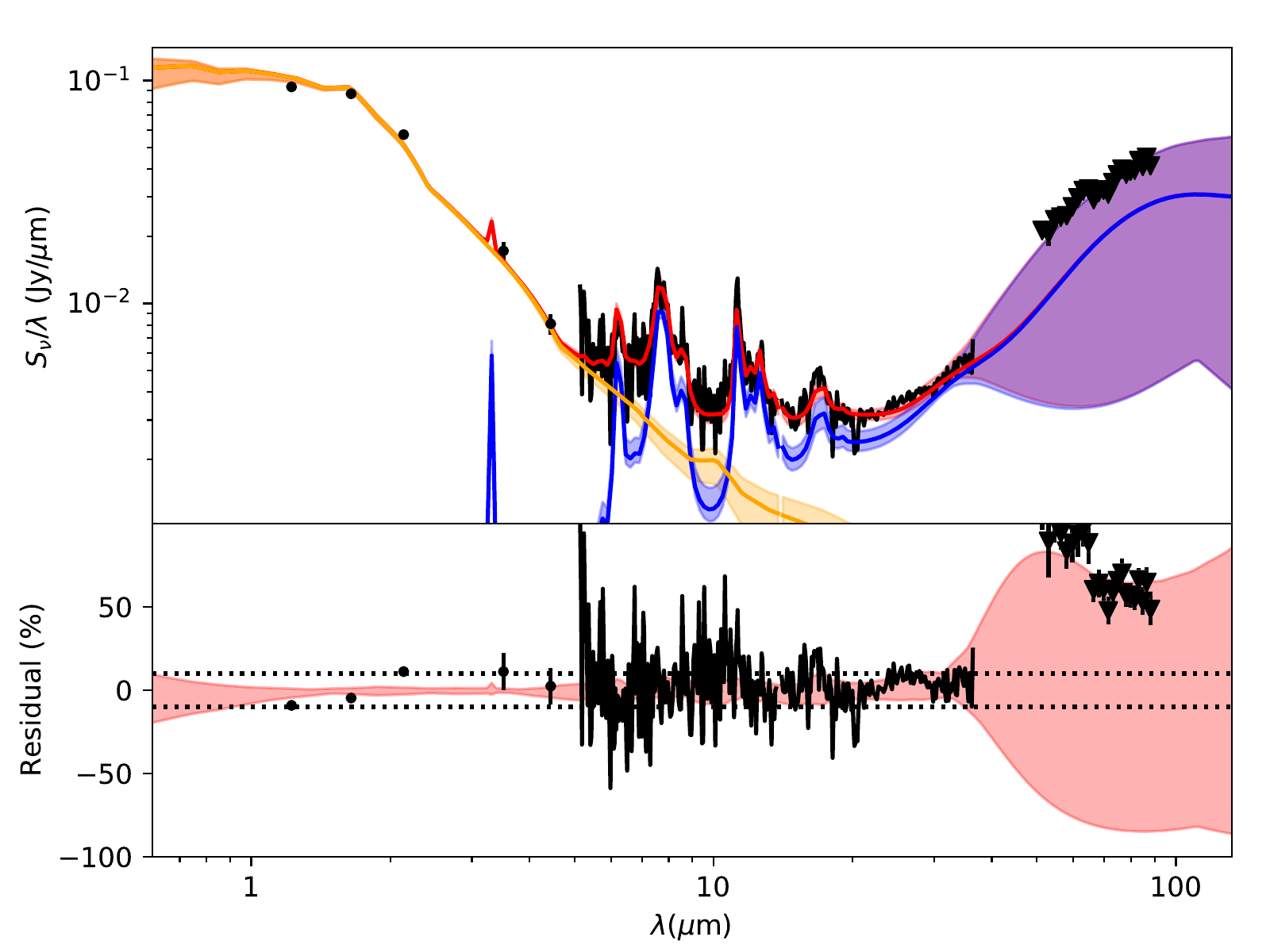}}
\caption{Plot of the infrared SED fit and its residual for NGC\,2639. Data are in black, with upper limits (inverted triangles) for larger apertures relative to Spitzer Infrared Spectrograph, IRS (in this case, Multiband Imaging Photometer, MIPS, SED mode). The starlight model is in orange, ISM model in blue, the sum is in red. Lines show the best fit, and the shaded region shows the range of realizations for that model component (or sum). The systematic uncertainty is expected to be about 10\%, indicated by dotted lines on the residuals plot. The statistical noise can exceed 10\% for the short wavelength end of the IRS spectrum.}
\label{figsed}
\end{figure}

We also estimated the star-formation rates from the Spitzer space telescope infrared fluxes using the clumpyDREAM code. The details of the fitting code itself is summarized in the appendix of \cite{sales2015}. It fits models simultaneously to infrared spectra and broadband SEDs. Since there was no evidence for a strong AGN component in the infrared, only an ISM model \citep{draine2007} and a simple stellar population from GRASIL \citep{silva1998} were adequate to fit the spectrum. Figure~\ref{figsed} shows the plot of the SED fit and its residual. The SFR turn out to be of the order of 1 M$_\odot$~yr$^{-1}$, lower than that estimated from the radio flux densities \citep[using Equation 28 in][]{condon2002}. This fact along with the collimated morphology of the lobes as well as high degree of linear polarization and an organised magnetic field structure, all point to an AGN origin for the radio lobes in NGC\,2639. 

\section{Discussion}
The discovery of an additional pair of radio lobes in our polarization-sensitive VLA observations at 5.5 GHz makes NGC\,2639 only the third known Seyfert galaxy after Mrk\,6 and NGC\,2992, with multiple pair of lobes on kpc scales aligned at different angles to each other. Multiple lobes have also been hinted at in the polarization image of NGC\,3079 in the CHANG-ES survey \citep{Irwin17}. We note that weak signatures of a north-south extension could be seen in previous images of NGC\,2639 in the literature using VLA and eMERLIN \citep[e.g.,][]{gallimore2006,baldi2018}. However, the high degree of polarization along with the ideal resolution of our EVLA image has helped us in discovering the second pair of lobes. The case of NGC\,2639 is similar to NGC\,2992 where an extra pair of lobes which was previously not revealed in total intensity due to contamination from the galaxy disk emission, was discovered in polarized emission by \cite{Irwin17}.

\begin{figure}
\centering{
\includegraphics[width=7.5cm,trim=0 80 0 150]{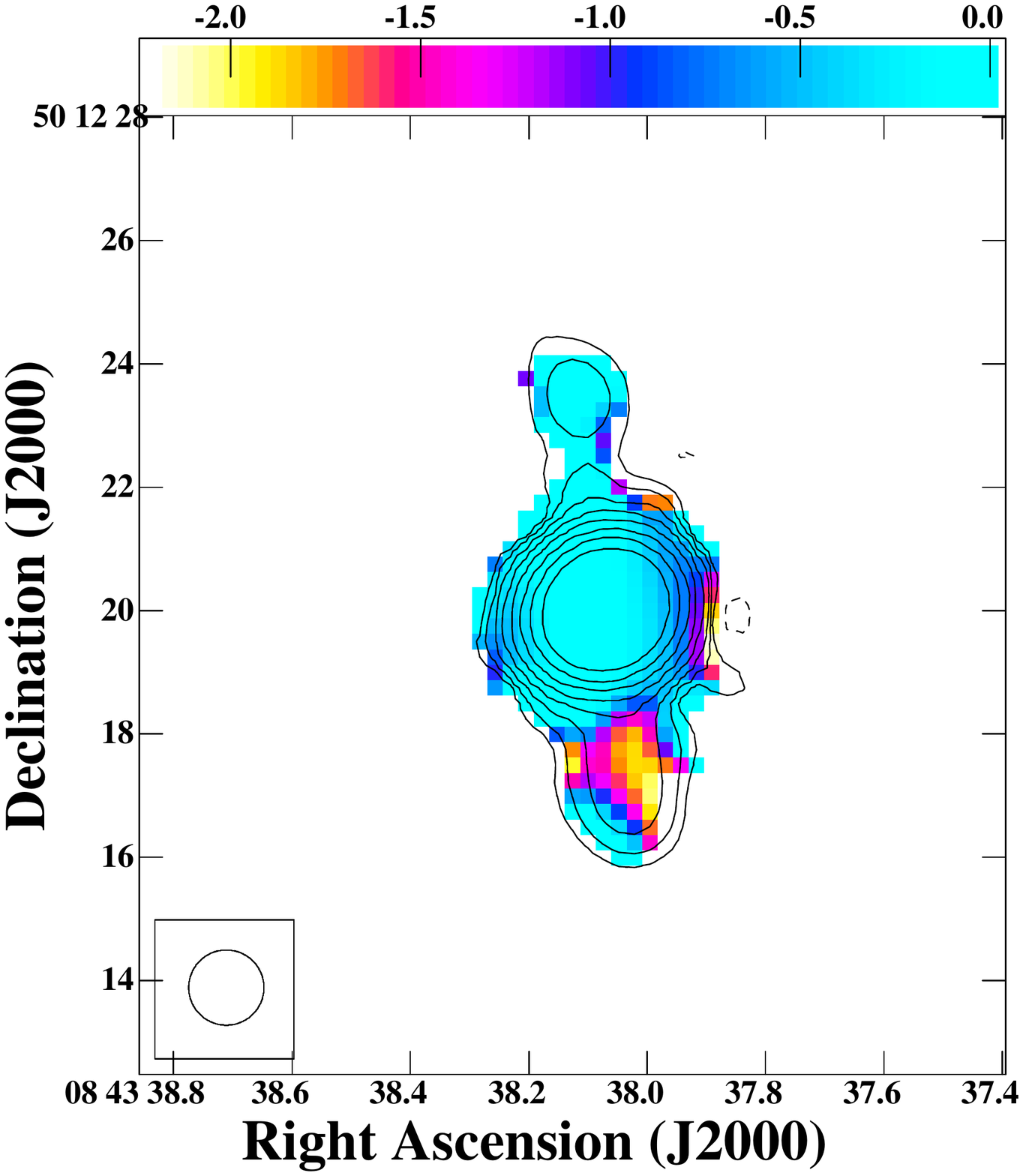}}
\caption{The in-band spectral index image of NGC\,2639 in color with 5 GHz radio contours superimposed. Contour levels: 50$\times$(1, 2, 4, 8, 16, 32, 64, 128, 256, 512) $\mu$Jy beam$^{-1}$. The beam shown in the lower left corner is of size $1.15\arcsec\times1.00\arcsec$ at PA =$-$4.9$\degr$.}
\label{fig4}
\end{figure}

\begin{figure}
\centering{
\includegraphics[width=7.5cm,trim=0 80 0 150]{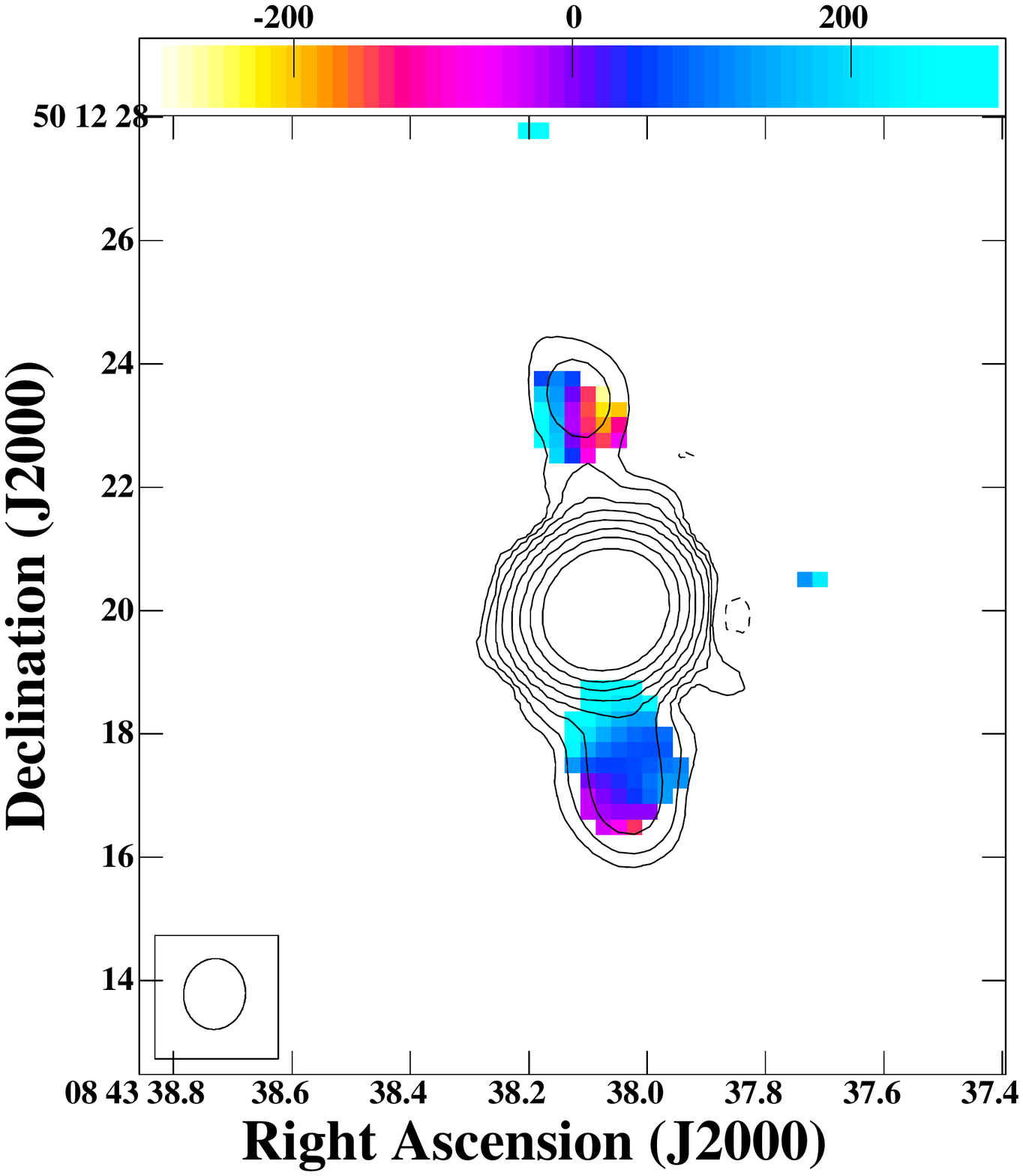}}
\caption{The in-band rotation measure image of NGC\,2639 in color with 5 GHz radio contours superimposed. Contour levels: 50$\times$(1, 2, 4, 8, 16, 32, 64, 128, 256, 512) $\mu$Jy beam$^{-1}$. The beam shown in the lower left corner is of size $1.15\arcsec\times1.00\arcsec$ at PA =$-$4.9$\degr$.}
\label{fig3}
\end{figure}

Several possibilities can lead to the observed double set of lobes in NGC\,2639, including (a) slow jet precession, (b) a pair of binary black hole at the center each launching independent jets, 
(c) recurrent jet activity at a misaligned angle. 

 The 5 GHz VLBI image of NGC\,2639 from \citet{Lal04} shows a curved parsec-scale jet bent towards the south. A curved jet could indicate jet precession, which in turn could arise from a warped accretion disk or the presence of binary black holes.
It is unlikely however, that the two sets of lobes is caused due to slow jet precession because of a lack of connection between these two jets as usually seen in the case of X or S-shaped sources. No continuity or connection is seen between the N-S and E-W lobes in the image of NGC\,2639 in Figure A1 from \cite{baldi2018}, where both the lobes can be seen simultaneously. We note however, that it is difficult to suggest what looks like diffuse extension in the north-south direction in their image to be a separate set of lobes in the absence of polarization data.

Alternately, the two episodes of AGN activity are a result of a minor gas-rich merger which created a new accretion disk in a direction driven by the angular momentum of the inflowing gas and which consequently created a jet in a different direction to the previous episode.
While the VLBI image does not indicate the presence of black hole binaries in NGC\,2639, sub-parsec-scale binaries cannot be ruled out with present day VLBI experiments. In addition, it is possible that currently both the black holes are not launching jets simultaneously, and hence both are not visible. The probability that the two sets of lobes result from a binary black hole pair with one set of lobes each is $<0.02\%$ if we consider the probability of binary black holes in optically-selected SDSS QSOs \citep[0.1\%;][]{Foreman09} and the probability of a Seyfert galaxy producing a kpc-scale radio lobes \citep[$\geq44$\%;][]{gallimore2006}. NGC\,2639 was one among the 10 Seyferts from our larger sample that was observed with the EVLA. NGC\,2992, which also shows two pairs of lobes is part of our sample as well. Hence, the probability of multiple set of lobes estimated from our sample is 20$\%$, which is much higher than the expected occurrence rate of binary black holes launching jets. Of course we note the caveat of our small sample size.
Given the relatively undisturbed morphology of the host galaxy, any merger that might have contributed a second supermassive black hole must have been minor. Since the black hole mass and bulge mass of a galaxy are correlated \citep{magorrian1998}, it is probable that the second black hole mass would be much smaller and hence less likely to produce its own radio jet. For instance, \citet{Baldi18} found that black hole masses needed to be $ > 10^6$~M$_\odot$ for producing radio jets. 
Hence, both the probability estimates and the undisturbed morphology of the host galaxy argues against the feasibility of the binary black hole model.

\cite{Irwin17} consider the scenario where the lobe misalignment occurs as a result of interaction with ambient medium, which deflects the radio lobes along the direction of steepest pressure gradient. Given that in NGC\,2639 the lobes are well confined within the galaxy itself, interaction with the ISM is inevitable. With additional estimates of parameters like the pressure exerted by the ISM on the lobes, which can be derived from X-ray measurements, useful constraints on the amount of jet bending can be obtained \citep{fiedler1984}. Such bending should lead to an alignment with the galaxy minor axis. The PAs of various components listed in Section~\ref{secresults} suggest that the east-west lobe is misaligned with the host galaxy minor axis by $\sim60\degr$, whereas the north-south lobe is misaligned only by $\sim40\degr$. The fact that the galaxy minor axis lies in between the current position angles of both the set of lobes, undermines this model. Any bending induced by ISM interaction should have been closer to $60\degr$ rather than $\gtrsim90\degr$ that is observed in NGC\,2639.

 Lastly, we consider the episodic activity model.
Recurrent activity which leads to the formation of lobes at different angles is a possibility that has been invoked to explain such structures in radio galaxies \citep{nandi2017}. A comparison of images at different resolutions can provide us a handle on the currently active lobes. The parsec-scale jet in \cite{Lal04} pointing towards the east in NGC\,2639 might have been indicative of the large-scale east-west lobe being the currently active jet. In addition to the jet bending seen in parsec scales \citep{Lal04}, the jet seen in the archival VLBA image presented in this paper which is at a lower resolution compared to \citet{Lal04} is aligned at an angle which is in-between both of the kpc scale lobes. It is therefore possible in principle, that the north-south lobes could be fed by the VLBI jet. Moreover, the relative total intensity of the eastern and western outflows inside the `core' (top right panel of Figure~\ref{fig2}) clearly shows the western one to be brighter and perhaps Doppler-boosted on $\sim$200 parsec-scales, whereas the eastern jet is the approaching one in the VLBI images. It is however possible that the asymmetry in brightness on the 100-parsec-scale is  result of asymmetric environments rather than Doppler-boosting. While the degree of polarization is comparable in the two lobes to the north and south (it is marginally higher in the northern lobe, which could be due to an intrinsically highly polarized jet/knot region), the fact that the southern lobe shows a larger polarized lobe extent is consistent with it being the approaching lobe \citep[e.g.,][]{Laing88}. The fact that all angles are measured in projection, must of course, be borne in mind.

From the image of NGC\,2639 presented in \cite{baldi2018}, it can be seen that the north-south lobes appear more diffuse, whereas the east-west lobes are more collimated which might be suggestive that the north-south lobes form the older set of lobes which had more time to expand adiabatically even after the switching off of the central engine. The spectral index values are also flatter in the `core' compared to that in the north-south lobes, which is also in accordance with north-south lobes forming the older epoch of emission. 
\cite{gallimore2006} quote a duty cycle of $\geq$ 44\%. An average duration of each of the outflows can be determined from the duty cycle which is defined as $\tau_{flow} =44\tau_{8}$ n$ ^{-1} $ Myr. Here $\tau _8$=1 is the typical Seyfert activity time-scale in 10$^8$ yrs and n is the number of outflows. For n=2 and assuming $\tau _8$=1, we derive an outflow duration of 22 Myr which is also comparable to the derived electron lifetime of the northern lobe ($\sim16$~Myr).

Could the restarted AGN activity in NGC\,2639 be a consequence of a minor galaxy merger? \citep[e.g.,][]{Capetti06}. The SDSS optical image however, does not show any tidal tail features or other outward signatures of a recent merger. The host galaxy does possess a bar and a stellar ring \citep[e.g.,][]{Marquez99} which could be indicative of morphological irregularities caused by a past gas-rich merger. It is also noteworthy that there has been a $\sim90\degr$ change in the jet direction between the two AGN activity episodes, similar to what is observed in Mrk\,6 by \citet{Kharb06}. Such rapid jet flips have been suggested to be a consequence of the presence or mergers of binary supermassive black holes \citep{Merritt02,kharb2017}. 

The lack of polarization from most of the `core' region is likely to be due to the presence of copious amounts of magneto-ionised gas belonging to gas in the broad-line region (BLR), the narrow-line region (NLR) or the central ISM of the host galaxy. Using a very high-resolution study of core-dominated AGN, \cite{codea1989} puts an upper limit of 200~rad~m$^{-2}$ on the integrated RM introduced by the magneto-ionic medium. The north-south lobes on the other hand show an r.m.s. RM value of $\sim$125~rad~m$^{-2}$. For an average magnetic field value of $\sim13\mu$G and a path length of $\sim$100~parsec, we obtain an electron density $n_e\sim0.1$~cm$^{-3}$. The volume averaged electron density typically seen in the warm interstellar medium (WIM) ranges from 0.01~cm$^{-3}$ to 0.1~cm$^{-3}$ \citep{gaensler2008}. On the other hand, \cite{kaakad2018} show that the electron densities in the NLR region varies from $\leq50-2000$~cm$^{-3}$. The observed RM in NGC\,2639 could therefore be due to WIM or the intercloud NLR gas. As \cite{codea1989} point out, the RM could still arise from NLR if the path-length is overestimated, or if the magnetic field is highly tangled leading to less apparent rotation of the electric vectors.

The magnetic field orientation is aligned with the lobe direction both in the north and the south, but there is a 30$\degr$ change in magnetic field orientation right at the end of the lobes. These could indicate a change in the flow direction of the radio emitting plasma, as in an S-shaped radio jet. 
 While the magnetic fields are aligned along the lobe axis in both the north and south lobes, the RM gradient in the northern lobe is suggestive of a large-scale helical or toroidal magnetic field \citep{Asada02,Kharb09}, whereas the southern lobe shows an RM gradient in the longitudinal direction, which points to a greater complexity in the magnetic field structure.

%

\section{Summary}
We report here the discovery of a new pair of north-south oriented radio lobes in the Seyfert galaxy NGC\,2639; this galaxy has been known to possess an east-west core-jet structure. The in-band RM image shows gradients in both the lobes indicative of organised magnetic field structures on kpc-scales. The magnetic field structure is aligned with the jet/lobe direction in both the lobes, but shows a change in orientation around the terminal point; an S-shaped radio outflow is implied. The presence of a twin pair of orthogonally aligned radio lobes in NGC\,2639 is highly reminiscent of the Seyfert galaxy Mrk\,6. Based on the settled optical morphology of the host galaxy, it is likely that a minor merger that did not disrupt the host galaxy structure is responsible for the observed features in NGC\,2639. This also explains the near 90$\degr$ change in the jet direction; the current jet direction being the result of a new accretion disk formed by the minor merger, whose direction was a result of the angular momentum of the inflowing merger gas. 
 Additional radio images at better resolution revealing more details of the north-south outflow along with the east-west lobes, can provide more definitive clues. For example, determining the ages of both the lobes simultaneously can give insights into the time elapsed between the stopping and restarting of the jets, which can in turn help in constraining various models \citep{dennett2002}. Furthermore, better resolution polarization images of the lobes can help identify any S-shaped outflow which is suggested by our current data. 


\bibliographystyle{mnras}
\setlength{\bibsep}{-2.2pt}

\bsp	
\label{lastpage}
\end{document}